\documentclass[12pt]{article}

\usepackage[margin=1.25in]{geometry}
\usepackage{graphicx}
\usepackage{cite}
\usepackage{array}
\usepackage{float}
\usepackage{hyperref}
\usepackage{stfloats}
\usepackage{xcolor}
\usepackage{color}
\usepackage{amsmath,amsfonts,amsthm}
\usepackage[ruled,vlined]{algorithm2e}
\usepackage{caption}
\usepackage{subcaption}
\usepackage{multirow}

\DeclareMathOperator*{\argmax}{arg\,max}
\DeclareMathOperator*{\argmin}{arg\,min}

\newtheorem{theorem}{Theorem}

\newtheorem{prop}[theorem]{Proposition}

\newcommand{\bbR}{{\mathbb R}}

\newcommand{\bfu}{{\mathbf u}}
\newcommand{\bfv}{{\mathbf v}}

\newcommand{\um}{\textcolor{black}}

\begin{document}

\title{Testing Rank of Incomplete Unimodal Matrices}

\author{Rui Zhang$^a$, Junting Chen$^b$, Yao Xie$^a$,\\ Alexander Shapiro$^a$, Urbashi Mitra$^c$}

\maketitle

\begin{abstract}
Several statistics-based detectors, based on unimodal matrix models, for determining the number of sources in a field are designed.  A new variance ratio statistic is proposed, and its asymptotic distribution is analyzed.  The variance ratio detector is shown to outperform the alternatives.  It is shown that further improvements are achievable via optimally selected rotations.  Numerical experiments demonstrate the performance gains of our detection methods over the baseline approach.

\end{abstract}


\section{Introduction}


Source localization based on received signal strength (RSS) \cite{meesookho2007energy} is challenged in environments where universal signal models are absent, such as underwater scenarios. 
Recently, a semi-parametric method based on sparsely spatially sampled RSS values and unimodal matrix factorizations has shown strong promise\cite{chen2019unimodality} for localizing multiple sources. However, \cite{chen2019unimodality} requires {\em a priori} knowledge of the number of sources.
In practice, such information is unknown and must be learned.
Herein, we design methods for estimating the number of sources when the signal propagation model is unknown save for the general property that the signal strength decreases with range.
While the rank of the observation matrix, in specific cases, can be indicative of the number of sources, this is not true in general.  For example, a single source can yield an observation matrix with rank two or more.  Alternatively,  if sources are co-linear,  the rank of the observation matrix can be less than the number of sources.

Prior art on source localization also required the number of sources (e.g., \cite{sheng2004maximum}).  If specific parametric models for signal strength exist, they can be leveraged to estimate the number of sources as in\cite{ottersten1993exact,viberg1994bayesian}.  Our goal, herein, is to provide methods for very general signal models such as unimodality.

This paper develops a statistical test on the rank of the sparse observation matrix to estimate the number of sources based on RSS measurements from different locations. Our proposed method does not require a parametric model for the signal, and we also assume the noise variance is unknown. We tackle the challenge of testing the rank of a low-rank matrix using incomplete observations while assuming the signal energy field is unimodal. The proposed method is based on extending the results in \cite{shapiro2018matrix, zhang2019statistical} to unimodal matrices. Moreover, we develop a new {\it variance ratio statistic} and the corresponding test procedure.  We further derive the asymptotic distribution of our proposed statistic, enabling the design of a 
threshold to control the false alarm rate. The proposed detection statistic is asymptotically efficient compared to an alternative estimator with an independent numerator and denominator. Furthermore, to improve our detector performance, we compute an optimal rotation to avoid the scenario of worst-case co-linearity of sources 
\cite{chen2019unimodality}.  Numerical experiments validate the theory and demonstrate the proposed procedure's good performance. 

\section{Problem Setup}
\label{sec:proSet}
Consider the problem of $K$-source localization.  Each source emits a signal that is unimodal:  maximum signal strength for a measurement made at the source location, with reduced signal strength for measurements made away from the source location, decaying with the distance from the source location.  For a single source, $k$, there is an $N \times N$  signal strength matrix $H^{(k)} \in \bbR^{N\times N}$. Let the singular value decomposition (SVD) of $H^{(k)}$ be
\[H^{(k)} = \sum_{i=1}^N \lambda_{k,i}\mathbf u_{k,i}\mathbf v_{k,i}^\top, ~k = 1, \ldots, K,\] 
where $\lambda_{k,1}\ge\lambda_{k,2},\dots,\ge\lambda_{k,N}$. A key result from \cite{chen2019unimodality} is that if the source signal is unimodal, then the singular vectors ($\bfu_{k,1}$ and $\bfv_{k,1}$) associated with the dominant singular value are also unimodal.  Thus, we approximate the signal matrix for source $k$ as,
\[H^{(k)} \approx \lambda_{k,1}\mathbf u_{k,1}\mathbf v_{k,1}^\top, ~k = 1, \ldots, K,\] 
Without loss of generality, we assume that $\bfu_{k,1}$ and $\bfv_{k,1}$ have positive components.



%
With our approximation in hand, 
the composite energy field is the superposition of the matrices of the $K$ sources, which can be written as, where the second double sum is due to contributions of the non-dominant singular values:
\begin{align*}
	H := \sum_{k=1}^K H^{(k)} &= \sum_{k=1}^K\lambda_{k,1}\bfu_{k,1}\bfv_{k,1}^\top +\sum_{k=1}^K\sum_{i=2}^N\lambda_{k,i}\bfu_{k,i}\bfv_{k,i}^\top \\
	&:= S+ U,
\end{align*}
where will approximate $H \approx S$. The first term is referred to as the {\em structured} component, comprised of the sum of $K$ rank-one matrices due to the dominant singular values, and $U$  is the {\em unstructured} component.

\begin{figure}[H]
	\centering
	\begin{subfigure}[t]{0.4\linewidth}
	    \includegraphics[width=\textwidth]{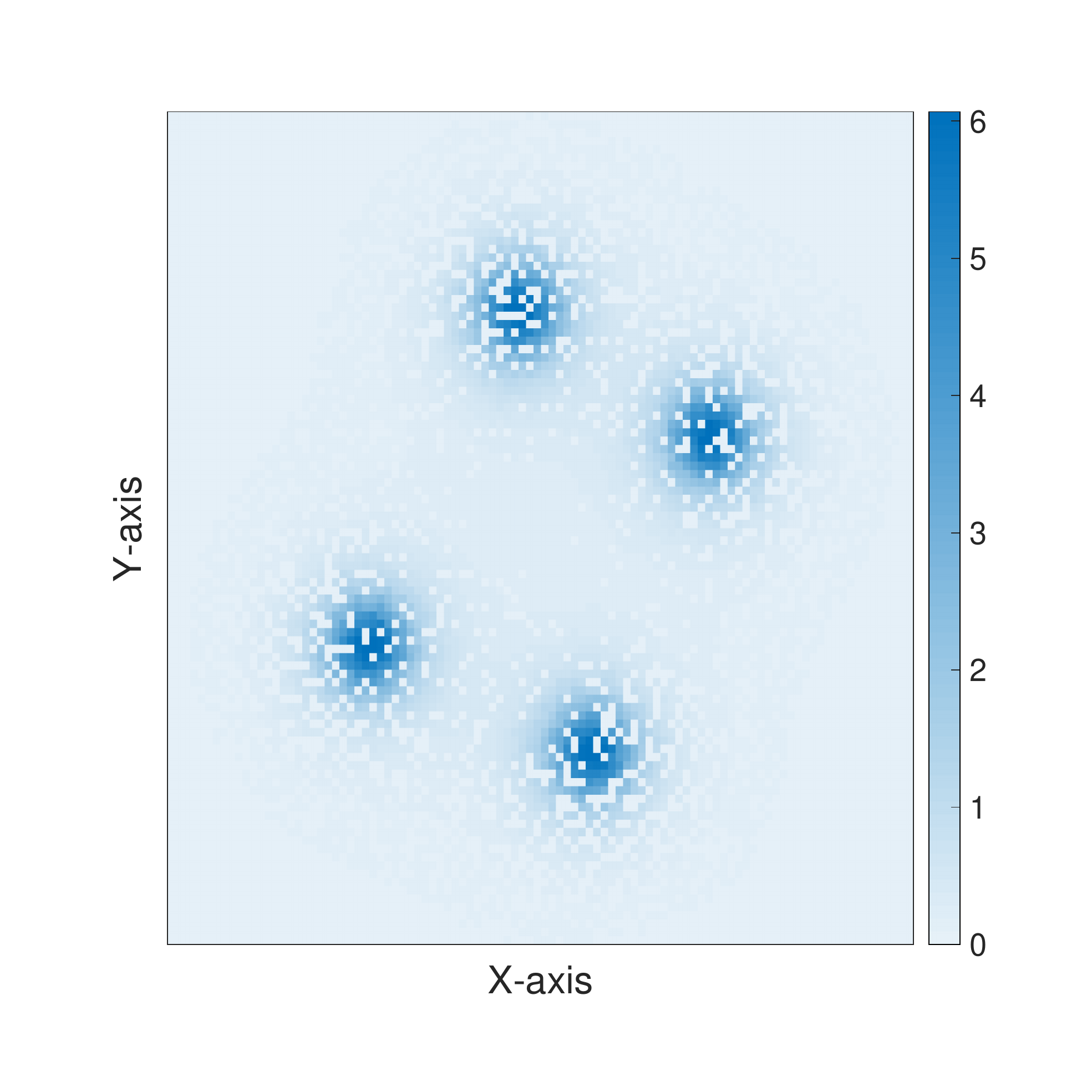} 
	\caption{}
	\label{fig-4sources}
	\end{subfigure}
	\begin{subfigure}[t]{0.4\linewidth}
    	\includegraphics[width =\textwidth]{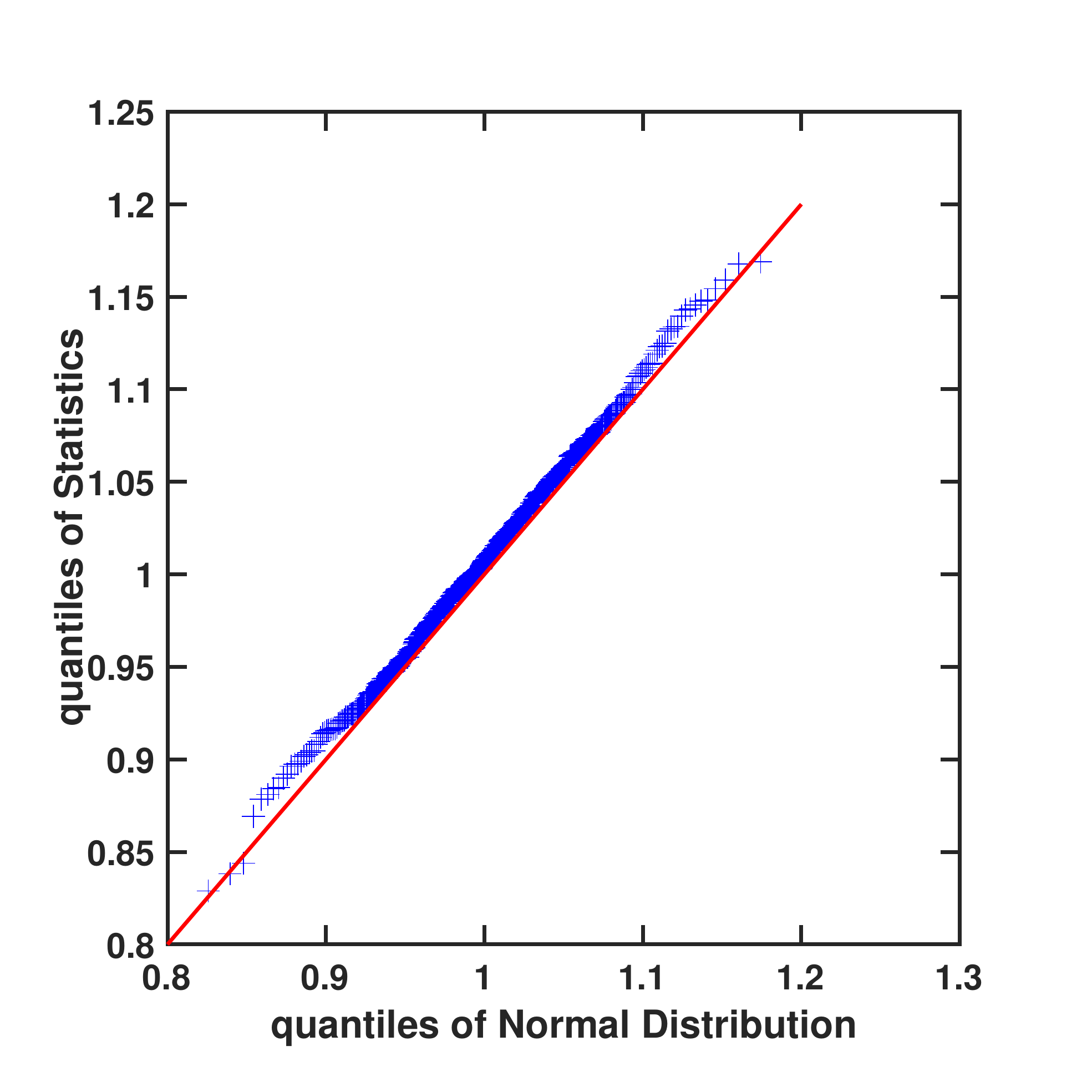}
    	\caption{}
	    \label{QQplot-thm}
	\end{subfigure}
	\caption{(a) Partially observed energy field with four sources; (b) QQ-plot: $\hat\sigma^2_2(\boldsymbol{Z})/\hat\sigma^2_1(\boldsymbol{Z})$ against quantiles of $\mathcal N(1, (c+2)/(2cL))$; the parameter values are $c =20$, $L = 150$.}
	\vspace{-0.1in}
\end{figure}

We assume access to a limited number of measurements at a subset of all possible locations.  The partially observed matrix $M$ is then given by,
\begin{align}\label{matrixM}
	M_{i,j} =  \begin{cases}
	S_{i,j} +\varepsilon_{i,j} &\forall (i,j)\in\Omega,\\
	0 &\forall(i,j)\notin\Omega,
	\end{cases}
\end{align} 
where $\Omega$ is the set of observation locations, and $\varepsilon$ is the measurement error on the $(i,j)$th observation \um{which encapsulates both measurement noise as well as the contributions due to the unstructured components}. Our goal is to identify the number of sources, $K$, given the partially observed matrix $M$. In Fig.~ \ref{fig-4sources}, we see examples of unimodal energy fields with four sources.

\vspace{-0.1in}

\vspace{-0.1in}
\section{Methods}

In this section, we propose two methods to identify the number of sources. As shown above, the rank of $S$ is closely related to the number of sources. The rank of $S$ is precisely equal to the number of sources if the unimodal property holds, and no rank-degeneracy is induced due to the summing of the contributions of the multiple sources, such as co-linearity. Therefore, determining the number of sources can be posed as finding a low-rank representation of $M$.

Our first method exploits the idea of rotation as proposed in \cite{chen2019unimodality} to mitigate the possibility of rank-degeneracy and thus improve performance. In the second strategy, we develop a heuristic method that considers multiple rotations and computes a statistic based on summing the dominant singular values obtained from each rotation.


\vspace{-0.1in}
\subsection{Method I: Variance ratio statistics with optimal rotation}

Herein, we adapt our prior statistical analysis in \cite{shapiro2018matrix} to the signal model herein Equation (\ref{matrixM}). Let $M, S\in\bbR^{n_1\times n_2}$, $r^* = {\rm rank}(S)$, and assume $\varepsilon_{i,j}\stackrel{i.i.d.}{\sim} N(0, \sigma^2) $, $\forall (i,j)\in\Omega$. 

In practice, the variance of the noise is often unknown and needs to be estimated. In \cite{zhang2019statistical}, a method to estimate the $\sigma^2$ and determine the rank heuristically is proposed.  In contrast, here,  we propose a statistic that does not depend on the scale of $\sigma^2$ and derive the asymptotic distribution. Therefore, we can provide a statistical hypothesis test for the rank.

To construct the statistic, we sub-sample from $\Omega$ to create a sequence of observations sets, such that $\Omega =\Omega_0\supset\Omega_1\supset\dots\supset \Omega_L$ 
and $|\Omega_i|-|\Omega_{i-1}| = c,\,\forall 1\leq i \leq L$, where $L$ is the number of  times we sub-sample and
$c$ is the number of observations we ``leave-out'' in between two consecutive sets. 
 
Let ${\rm SSE}_l$ denote the 
residual of the matrix completion solution at rank-$r$: 
\begin{equation}\label{subpro}
	{\rm SSE}_l = \min_{Y:{\rm rank}(Y)=r}\sum_{(i,j)\in \Omega_l}\left(M_{ij}-Y_{ij}\right)^2.
\end{equation}
According to Proposition IV.3 in \cite{shapiro2018matrix}, if $r = r^*$, as $\sigma^2 \rightarrow 0$,
\begin{align*}
  	&\frac{Z_l}{\sigma^2} = \frac{{\rm SSE}_{l-1} -{\rm SSE}_l}{\sigma^2} \rightarrow \chi^2(c), \,\,\,\forall 1\leq l \leq L.
\end{align*}
Let
\begin{equation}\label{def-sigma1}
 	\hat\sigma^2_1(\boldsymbol{Z}) = \frac{\sum_{l=1}^L Z_l}{cL},
\end{equation}and
\begin{equation}\label{def-sigma2}
 	\hat\sigma^2_2(\boldsymbol{Z})  = \sqrt{\frac{\sum_{l=1}^L (Z_l-\bar Z)^2}{2cL}},
\end{equation}
where $\boldsymbol{Z} = (Z_1,\dots,Z_L)$ and $\bar Z = (\sum_{l=1}^L Z_l) / L$. \um{We essentially use the method of moments to estimate the $\sigma^2_i$: the first moment in Equation (\ref{def-sigma1}) for $\hat{\sigma}_1^2$ and the second moment in Equation (\ref{def-sigma2}) for $\hat{\sigma}_2^2$.}

From the law of large numbers, we have that
 $\hat\sigma_1^2\rightarrow \sigma^2$ and $\hat\sigma_2^2\rightarrow \sigma^2$,  as $L\rightarrow\infty$. From our numerical results, we observe that when the matrix approximation problem in Equation (\ref{subpro}) is solved with the estimated rank smaller than the true rank, we tend to have $\hat\sigma^2_2 > \hat\sigma^2_1$. Therefore, we propose to use $\hat\sigma^2_2/\hat\sigma^2_1$ as the test statistic. In the experiments, we used singular-value-thresholding algorithms \cite{mazumder2010spectral}. However, the results are independent of the choice of the algorithm as long as it provides a good approximation for (\ref{subpro}). Fig. \ref{QQplot-thm} confirms that our asymptotic analysis in Theorem 1 matches well with the simulated result.
\begin{theorem}[Asymptotic distribution of variance ratio test statistic]\label{thm-statsAsym}
	Suppose $Z_1,\dots, Z_L$ are $i.i.d.$ random variables, and $Z_1/\sigma^2\sim \chi^2(c)$. Then, $\hat\sigma^2_1$ and $\hat\sigma^2_2$ defined in (\ref{def-sigma1}) and (\ref{def-sigma2}), respectively, satisfy
	\[
		\sqrt{\frac{cL}{2}}\left(\frac{\hat\sigma^2_2(\boldsymbol{Z})}{\hat\sigma^2_1(\boldsymbol{Z})} - 1\right) \stackrel{d}{\rightarrow} \mathcal N\left(0, \frac{c+2}{4}\right),
		\]
	as $L\rightarrow \infty$. $\stackrel{d}{\rightarrow}$ denotes convergence in distribution.
\end{theorem}


Since the parameters of sub-sampling, $c$ and $L$, which are defined above (\ref{subpro}), are known in practice, we can control the type-I error by choosing the threshold by Theorem \ref{thm-statsAsym}. Notice that $\hat\sigma^2_1$ and $\hat\sigma^2_2$ are dependent.  \um{Theorem \ref{thm-statsAsym} suggests a rank-estimation strategy: for each assumed rank $r$, we form the statistics and select the smallest $r$  such that the statistic fails to reject the null hypothesis.} Details are provided in Algorithm \ref{alg-1}.

\begin{algorithm}[H]
\label{alg-1}
\SetAlgoLined
\textbf{Input}: $M$, $\Omega$, $r_{\rm max}$, $c$, $L$, $b$.\\
 $\hat r, r = 0$\;
 \While{$r+1<r_{\rm max}$ }{
 $r = r+1$\;
  compute $\hat\sigma^2_1$ and $\hat\sigma^2_2$ as in eq. (\ref{subpro}), (\ref{def-sigma1}) and (\ref{def-sigma2})\;
  \If{${\hat\sigma^2_2}/{\hat\sigma^2_1}<b$}{$\hat r = r$\;
   }
 }
 \KwResult{$\hat r$}
 \caption{Rank detection with variance ratio. }
\end{algorithm}

\subsubsection{Relative efficiency}

Notice that $\hat\sigma^2_1(\boldsymbol{Z})$ and $\hat\sigma^2_2(\boldsymbol{Z})$ are dependent. One can construct two independent estimates by performing more sub-sampling, i.e. $\Omega =\Omega_0\supset\Omega_1\supset\dots\supset \Omega_{2L}$. Let $\boldsymbol{Z}_{1:L}$ denote the first half of $(Z_1, \dots, Z_{2L})$ and $\boldsymbol{Z}_{L+1:2L}$ denote the second half. Then $\hat\sigma^2_1(\boldsymbol{Z}_{1:L})$ and $\hat\sigma^2_2(\boldsymbol{Z}_{L+1:2L})$ are independent. 
The following proposition shows that our proposed variance ratio statistic $\hat\sigma^2_2(\boldsymbol{Z}_{1:L}) /\hat\sigma^2_1(\boldsymbol{Z}_{1:L}) $ is asymptotically more efficient than $\hat\sigma^2_2(\boldsymbol{Z}_{L+1:2L}) /\hat\sigma^2_1(\boldsymbol{Z}_{1:L}) $, i.e. the asymptotic variance of our proposed statistic is smaller.
\begin{prop}[Alternative variance ratio statistic]\label{thm-varRed}
	Suppose $\boldsymbol{Z} = (Z_1,\dots, Z_L)$ and $\boldsymbol{X}=(X_1,\dots, X_L)$, where $Z_i$ and $X_i$ are $i.i.d.$ random variables, and 
	\[\frac{Z_1}{\sigma^2} \stackrel{d}{=} \frac{X_1}{\sigma^2}\sim \chi^2(c),\] 
	 $\hat\sigma^2_1$ and $\hat\sigma^2_2$ are constructed as in (\ref{def-sigma1}) and (\ref{def-sigma2}),
 respectively. Then as $L\rightarrow\infty$, we have
	\begin{align*}
		\sqrt{\frac{cL}{2}}\left(\frac{\hat\sigma^2_2(\boldsymbol{Z})}
		{\hat\sigma^2_1(\boldsymbol{X})} - 1\right) \stackrel{d}{\rightarrow} \mathcal N\left(0, \frac{c+10}{4}\right).
	\end{align*}
\end{prop}
Comparing Theorem \ref{thm-statsAsym} with Proposition \ref{thm-varRed}, we can conclude that $\hat\sigma^2_2(\boldsymbol{Z})/\hat\sigma^2_1(\boldsymbol{Z})$ is asymptotically more efficient than 
	$\hat\sigma^2_2(\boldsymbol{Z})/\hat\sigma^2_1(\boldsymbol{X})$.   In practice, we only have a limited number of observations and can not sub-sample many times. 
With Theorem \ref{thm-statsAsym} and Proposition \ref{thm-varRed}, we have shown that we can construct a better statistic, with fewer samples, and attain improved performance by exploiting the dependence of the statistics in the numerator and estimator of our test.


\subsubsection{Rotation}
In \cite{chen2019unimodality}, it is observed that colinearity of sources can lead to rank degeneracy, and thus the number of sources may be larger than the rank of $S$.  It was further shown in \cite{chen2019unimodality}, that rotations can mitigate this issue and improve performance.  Thus, herein we consider multiple rotations of various degrees, and $E$ is a permutation matrix if and only if each row and column contains a single non-zero component of value one. 
Rotating a matrix $A\in \bbR ^{N\times N}$ with degree $\theta$ can be achieved in the following way:
$
A(\theta) = {\rm vec}^{-1}(E_\theta {\rm vec}(A)),$
where ${\rm vec}(A)$ is the vectorization of $A$ which stack columns of $A$ in a vector, ${\rm vec}^{-1}$ is the inverse function of ${\rm vec}$, and $E_\theta\in \bbR ^{N^2\times N^2}$ is the permutation matrix corresponding the degree $\theta$.

As stated previously, to avoid rank-degeneracy when applying the statistic in Theorem \ref{thm-statsAsym}, we need to look for a rotation such that most of the sources are not co-linear. We employ the following strategy as suggested in \cite{chen2019unimodality} to determine the optimal rotation, 
\begin{equation}\label{thetaOpt}
\theta_{\rm opt} = \argmin_\theta 
\frac{\lambda^2_1(H(\theta))}{\sum_{k=1}^N\lambda_k^2(H(\theta))},
\end{equation} 

\begin{figure}[h!]
	\centering
	\includegraphics[width = 0.35\linewidth]{F-test-energy-field2-min_H0_min.pdf}
	\includegraphics[width = 0.35\linewidth]{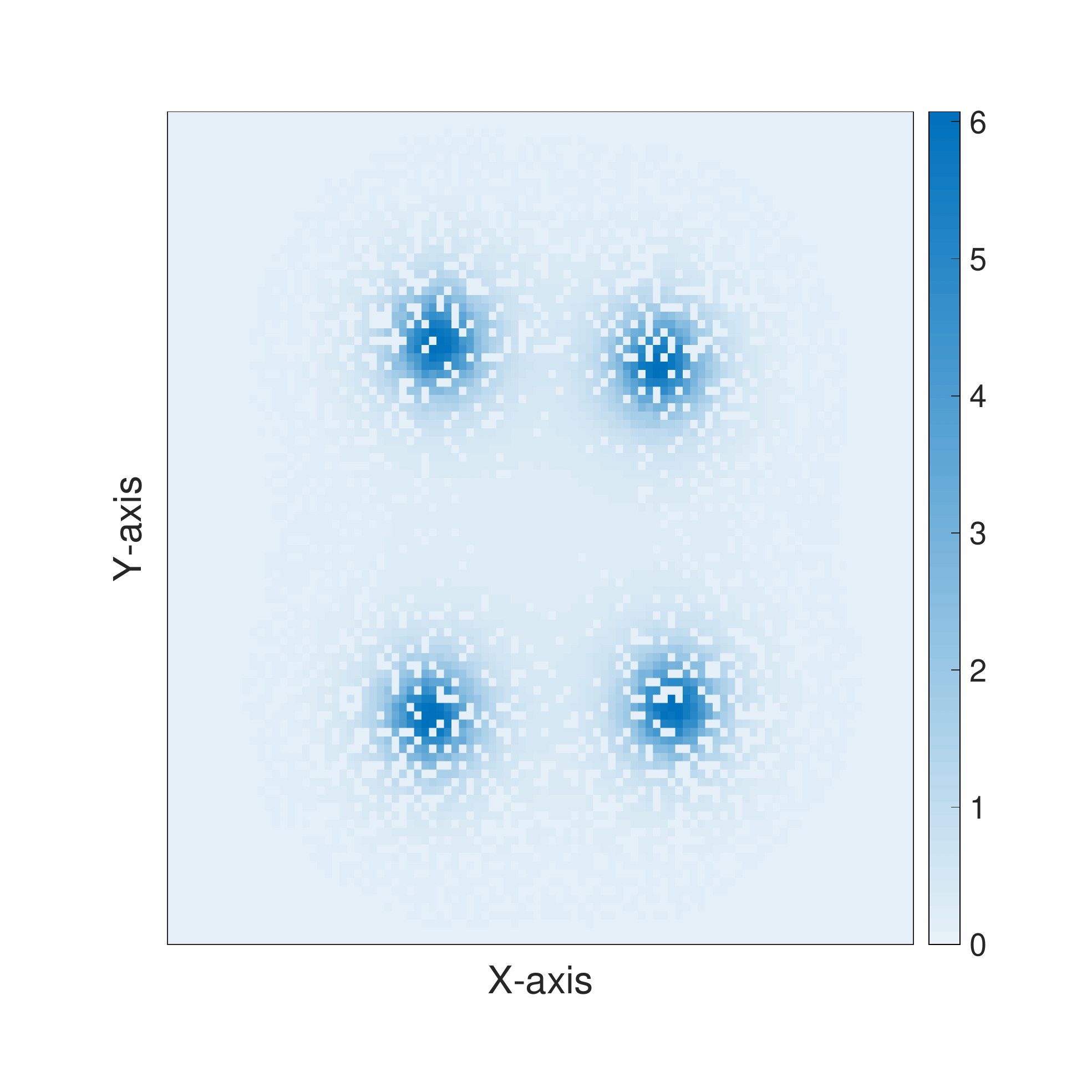}
	\caption{Left: Rotation with degree $\theta_{\rm opt}$. 
	Right: Rotation with degree $\theta_{\rm max}$.} 
	\label{rotationMinMax}
	\vspace{-0.2in}
\end{figure}

Numerical results suggest that the optimal rotation achieves the goal of avoiding rank degeneracy with high probability; furthermore, theu se of $\theta_{\rm opt}$ further improves the performance of our detector (see Theorem \ref{thm-statsAsym}). 
As a validating example, we examine the effect of rotatint the partially observed matrix by $\theta_{\rm opt}$ in (\ref{thetaOpt}), the optimal degree. 
Define $\theta_{\rm max} = \argmax_\theta \rho(\theta)$. Figure \ref{rotationMinMax} shows the
matrix after rotation with degree $\theta_{\rm opt}$ and $\theta_{\rm max}$, respectively. We can see rotation with $\theta_{\rm opt}$ avoids the alignment of
the sources; whereas rotating with $\theta_{\rm max}$ actually aligns the sources leading to rank degeneracy.

\subsection{Method II: Averaging effects of rotations}

\begin{figure}[h!]
	\centering
		\centering
		\includegraphics[width =0.35\linewidth]{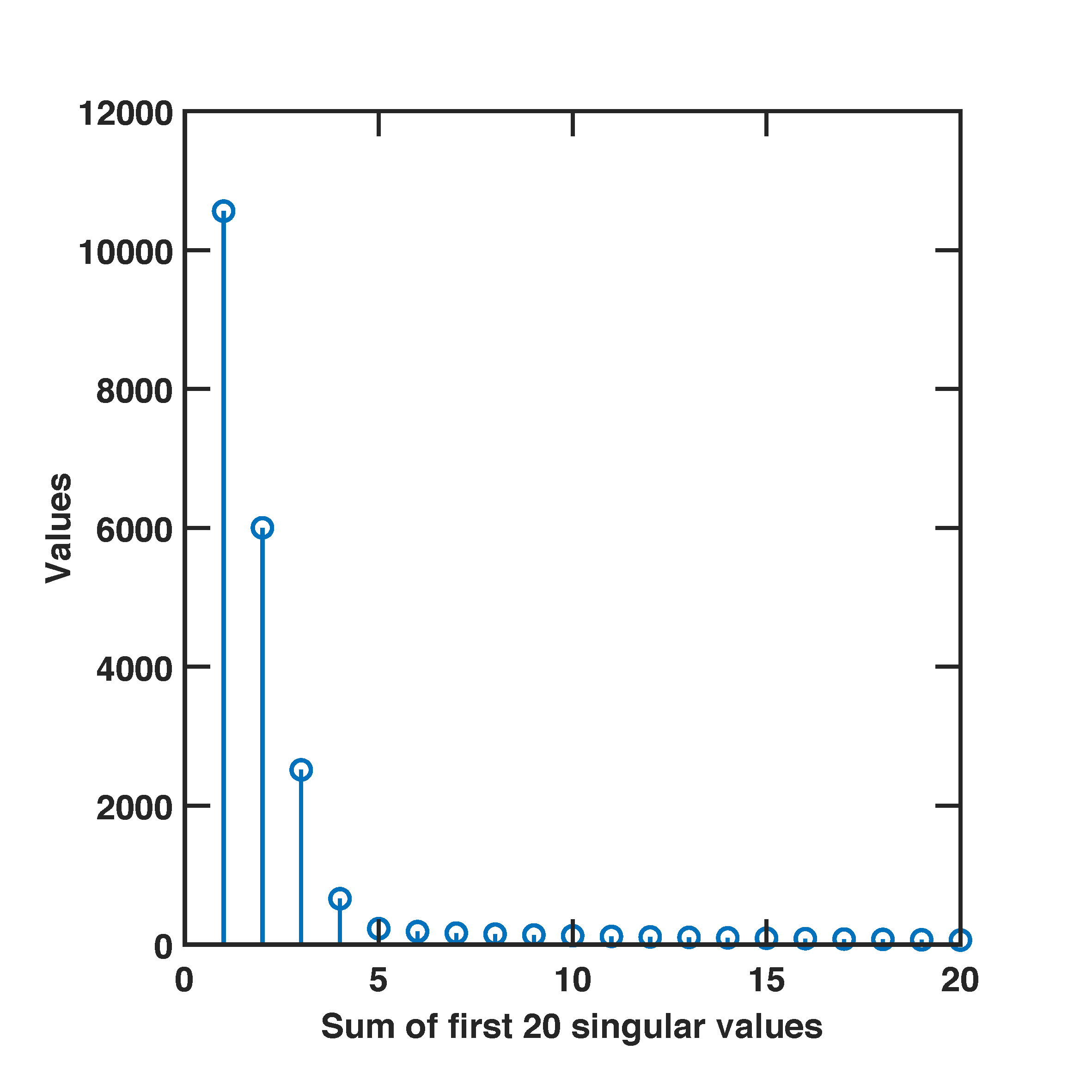}
		\includegraphics[width =0.35\linewidth]{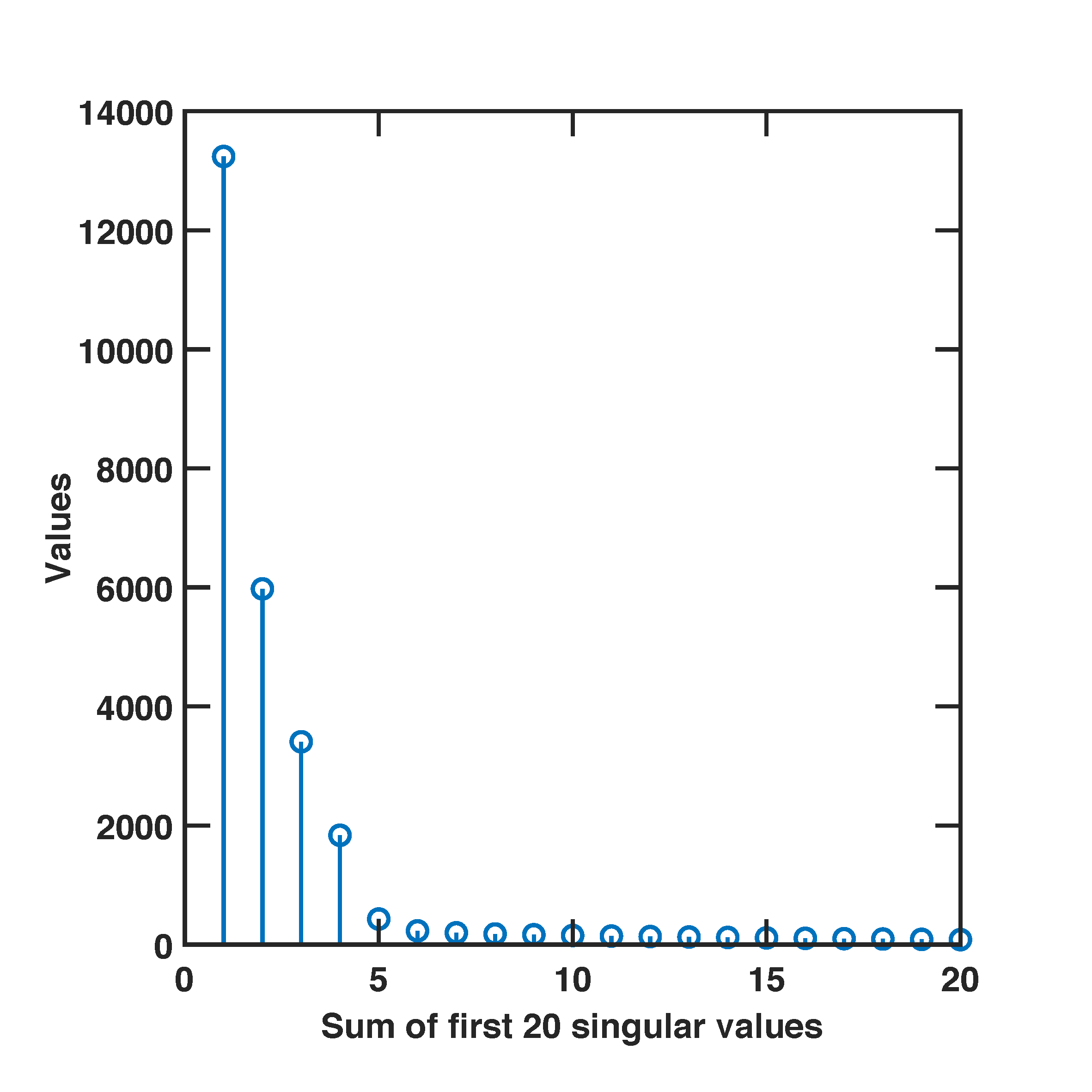}
		\caption{Sum of first 20 singular values of 100 rotation degrees. Left: $K = 3$. Right: $K = 4$.}
		\label{nn100K3-4}
		\vspace{-0.1in}
  \end{figure}

Instead of looking for an optimal rotation, we can rotate the matrix with multiple different degrees. For each rotation, matrix completion via nuclear norm minimization is conducted and the first few dominant singular values are computed. The resultant singular values are averaged over the multiple rotations.  This averaging reduces the effect of degeneracy by reducing the effects of ``bad degrees.''  This is a heuristic approach that can be applied even when the statistical conditions of the previous approaches do not hold.  Fig. \ref{nn100K3-4} shows
the result for 100 rotation degrees. We clearly see that that there are $K$ dominant singular values for $K$ sources, where $K=3,4$ respectively.



\begin{algorithm}[!h]
\SetAlgoLined
\textbf{Input}: $M$, $\Omega$, $n$, $D$, $b$, rotation degrees ($\theta_1,\dots, \theta_D$).\\
 \While{$i\leq D$ }{
 rotation $M$ with $\theta_i$ degree\;
 complete $M$ with nuclear-norm minimization\;
 compute first $n$ singular values: $\lambda_{i,1}, \lambda_{i,2},\dots, \lambda_{i,n}$\;
 $i = i +1$
 }
 \KwResult{$\hat r = \min\Big\{r, \frac{\sum_{j=1}^{D}\sum_{i=1}^r \lambda_{j,i}}{\sum_{j=1}^{D}\sum_{i=1}^n \lambda_{j,i}}>b\Big\}$}
 \caption{Rank detection with averaging rotations.}
 \label{alg-2}
\end{algorithm}

\vspace{-0.2in}

\section{Experiments}
\subsection{Theory validation for exact low rank matrix}

\begin{figure}[h!]
	\centering
	\includegraphics[width = 0.5\textwidth]{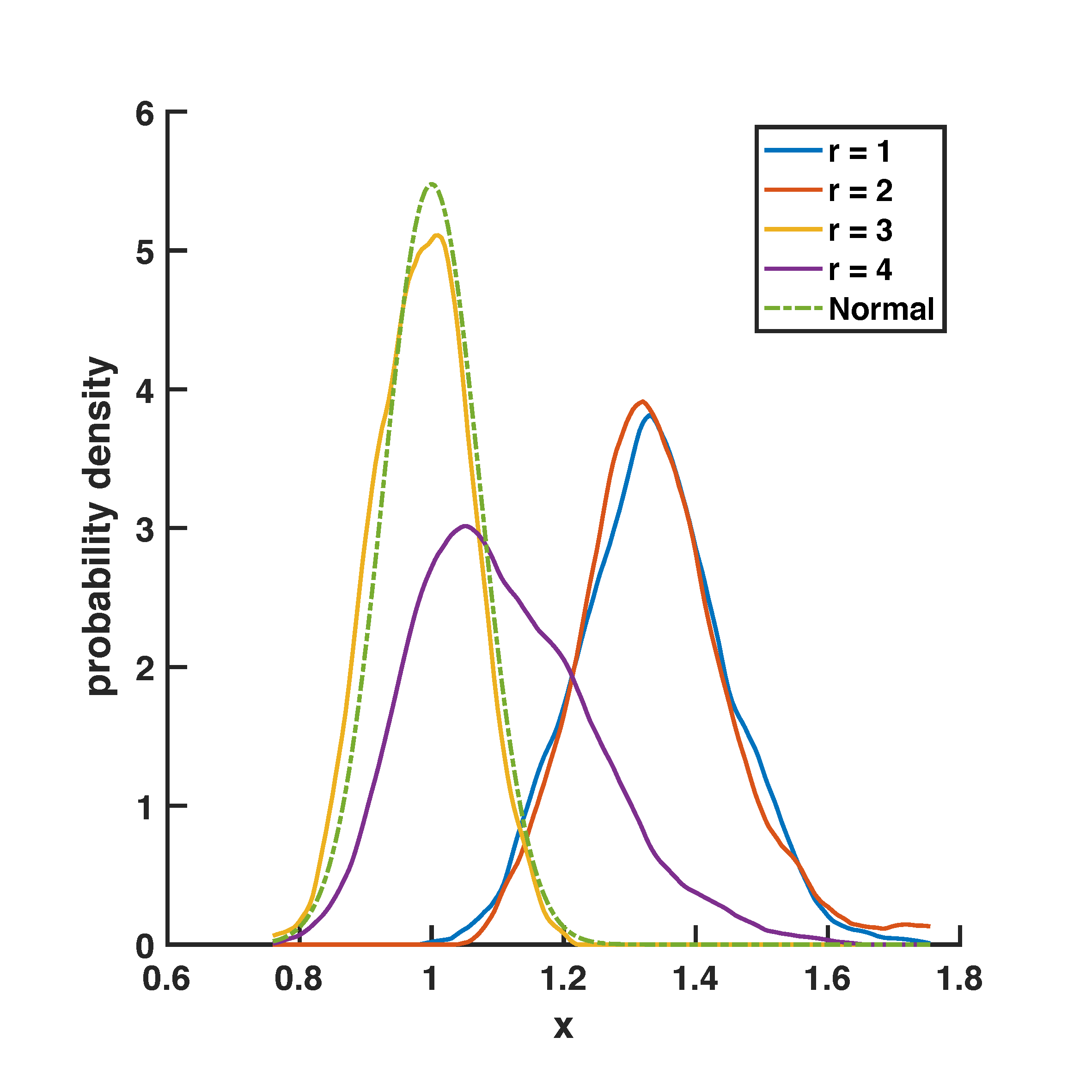}
	\caption{A low-rank matrix with $r^* = 3$: Solid lines: empirical density estimation at each rank with 200 experiments. Dash line: probability density
	function of $\mathcal N(1, 0.005)$, which is the asymptotic approximation by Theorem \ref{thm-statsAsym}. Clearly, it is a good fit when we compute our proposed statistics with the true rank.}
	\label{ELRM}
	\vspace{-0.2in}
\end{figure}

In this experiment, we show the distribution of our proposed statistics, when the data comes from the 
low rank matrix model in \cite{zhang2019statistical}, i.e. $M_{i,j} = Y_{i,j}+\epsilon_{i,j}$, $\forall (i,j)\in\Omega$ 
and $Y$ is a low rank matrix. We performed 200 experiments. In each experiment, a rank-3 matrix $Y\in\bbR^{100\times 100}$ is 
generated and $\Omega$ is uniformly randomly chosen, s.t. $|\Omega| = 7500$. $\epsilon_{i,j}\sim\mathcal N(0,1)$. 
Then, we compute the proposed statistics, under the assumption that the rank varies from to $1,\dots, 4$, with $c=30$ and $L = 100$. In Fig. \ref{ELRM}, we show the
empirical density estimates of our proposed statistics for each rank. When $r = 3$, which is the actual rank,
by Theorem \ref{thm-statsAsym}, our proposed statistics can be approximated by a normal random variable with mean  $1$, and variance equal to $0.005$. Fig. \ref{ELRM} shows a good fit for the empirical density corresponding to the true rank.
 For the rank less than the true rank, the density is well separated from the density
with true rank. However, when the rank is larger than the true rank, it is not. Therefore, by choosing the smallest rank such that the p-value of proposed statistic is larger than some threshold, we can find
the true rank with high probability.

\vspace{-0.1in}

\vspace{-0.1in}
\subsection{Determining number of sources in energy field}\label{energyField}
\begin{figure}[h!]
\vspace{-.1in}
	\centering
	\includegraphics[width = 0.35\linewidth]{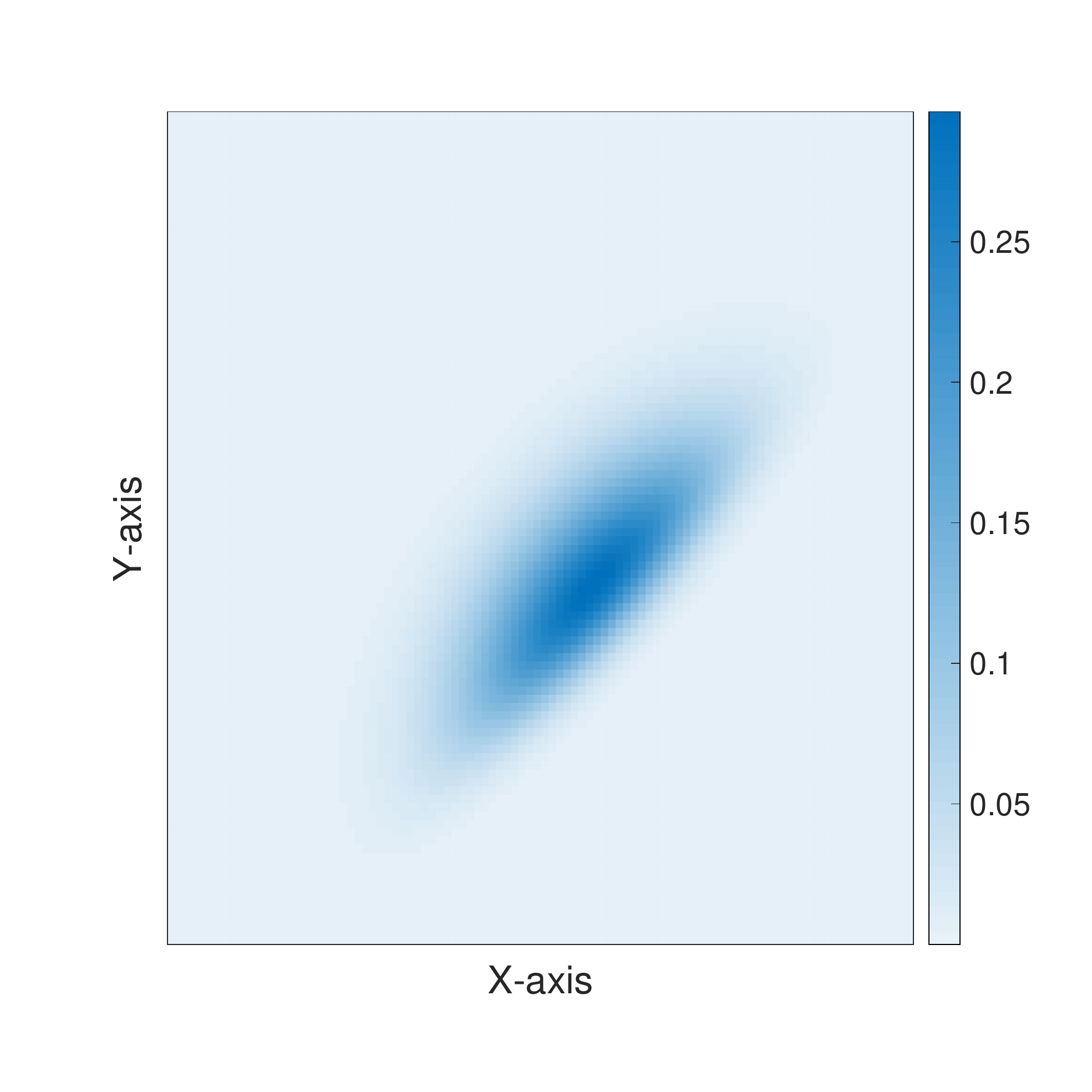}
	\includegraphics[width = 0.35\linewidth]{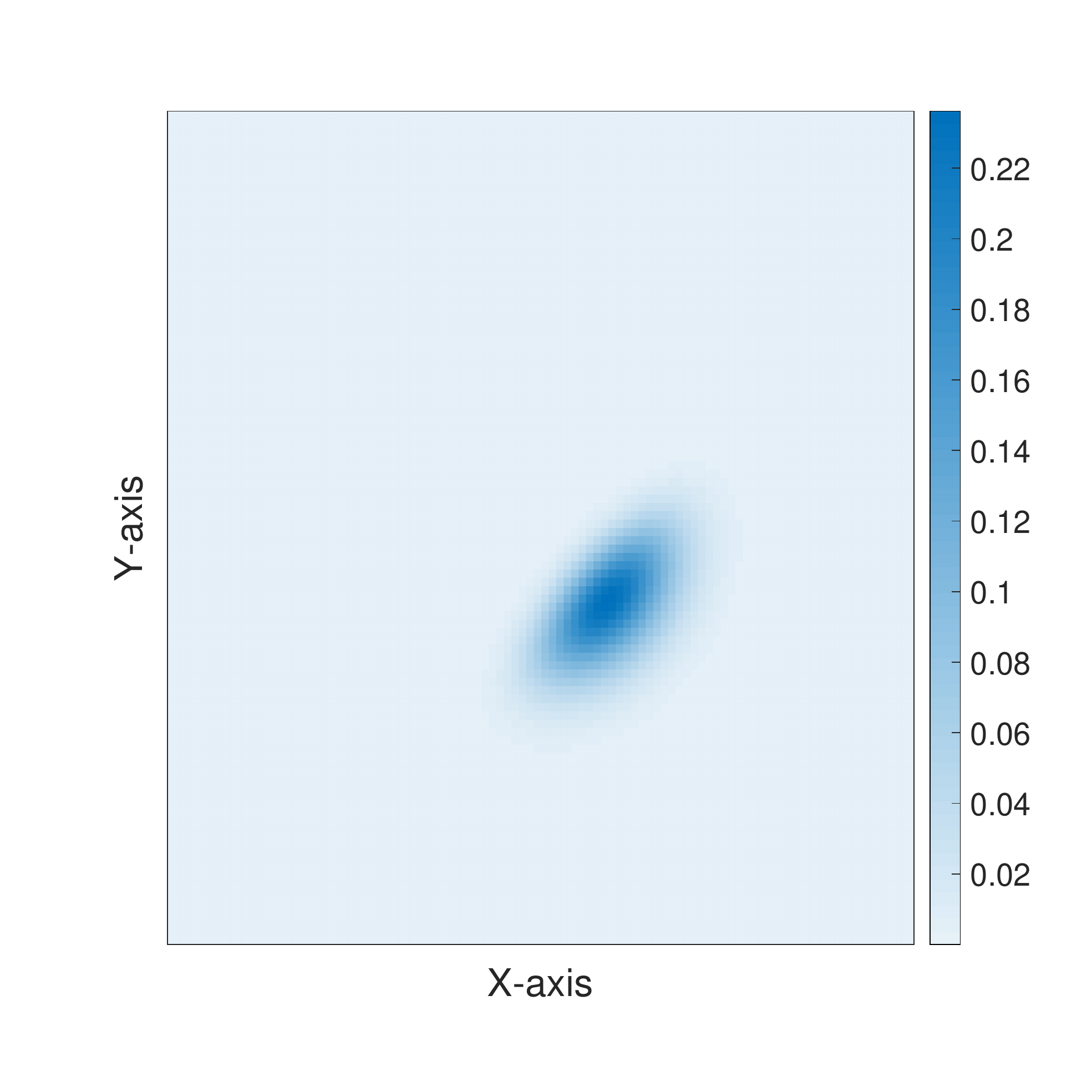}
	\caption{Skewed energy source modeled by bivariate skew-normal distribution. Left: $\delta_1 = \delta_2 = 0.8$, $\omega = -0.5$.
	Right: $\delta_1 = \delta_2 = 0.35$, $\omega = -0.25$.}
	\label{skewExample}
	\vspace{-0.1in}
\end{figure}

We next examine the efficacy of our method for detecting the number of sources via 
the confusion matrix and the $F_1$ score. Given multiple classes, we use the average $F_1$ score in \cite{opitz2019macro}, which is defined as follows: 
\[
  P_i = \frac{s_{ii}}{\sum_{x =1}^4s_{ix}}, R_i = \frac{s_{ii}}{\sum_{x=1}^4s_{xi}}, F_1 = \frac{1}{2}\sum_{i=2}^3\frac{2P_iR_i}{P_i+R_i},
\]
where $i$ is the estimated number of sources, $j$ is the true value and $s_{ij}$  denotes the number of occurences wherein the method estimates $j$ sources given $i$.
We consider both isotropic and skew  energy sources. 
\begin{itemize}
    \item {\it Isotropic sources:} For isotropic sources, the sampled energy is purely range-dependent and thus independent of the rotation of the coordinator.
In this experiment, we generate the energy field as the following. $af_{dB} = 0.11\times f^2/(1+f^2) + 44 \times f^2/(4100+f^2)
+2.75 \times 10^{-4}\times f^2 + 0.003$ with $f = 5$. $a(\mathbf x) = d(\mathbf x)^\alpha\times 10^{-af_{\rm dB}/10\times d(\mathbf x)}$, 
$P(\mathbf x) = p/(a(\mathbf x) + 1)$, where $\alpha = 3$ and $p = 6$. The field is square with diameter $15$ km. 
\item {\it Skew sources:} To generate the energy field with skew sources, we use the bivariate skew-normal distribution. According to Section 3 of \cite{azzalini1996multivariate}, $\delta_1$, $\delta_2$ and $\omega$ are the parameters to control the skewness. Figure \ref{skewExample} displays examples of skew sources. 
 In all experiments, $\delta_1$, $\delta_2$and $\omega$ are generated uniformly from $[-0.25, 0.25]$ for each energy source. 
\end{itemize}

\noindent

\noindent

All the experiments are repeated 200 times each for the number of sources $K^* = 2, 3$. The standard variance of ambient noise is $0.01$. The number of sensors is 4500, and we map them onto a matrix $M\in\bbR^{100\times 100}$. Notice that with the mapping, $|\Omega|\leq 4500$. Our proposed methods are compared with a baseline method.
\begin{itemize}
    \item 
{\it Baseline method: Zero filling.}
We apply singular value decomposition to the observation matrix $M$ in eq.(\ref{matrixM}) and determine the number of source by thresholding the percentage of leading singular values, i.e., given threshold $b$, the estimated number of source of baseline method, $\hat r_{BL} = \min \big\{r, \sum_{i=1}^r \lambda_i/\sum_{i=1}^{100} \lambda_i>b\big\}$.

\item
{\it Method I: Variance ratio statistic with optimal rotation.}
In each experiment, we find the optimal rotation first, and then apply Algorithm \ref{alg-1}
 with $r_{\max} = 4$, $c = 2$ and $L = 750$.
 
 \item
{\it Method II: Averaging effect of rotation.}
In each experiment, we apply Algorithm \ref{alg-2} with $n=20$, $D = 20$ and $\theta_i = (i-1)\pi/(2D)$, $\forall i =1,\dots, 20$. 
\end{itemize}
Results for isotropic and skew sources are provided in 
  Table \ref{table_ac1}.
The roman numerals indicate: I - variance ratio statistic with optimal rotation; II - averaging over multiple rotations; and III - zero padding, matrix completion and thresholding, the baseline approach.

For the isotropic sources, the methods I, II, and baseline thresholds are 2.25, 0.8, and 0.42, respectively. For the skew sources, the thresholds of the Methods I, II, and baseline are 1.57, 0.82, and 0.575, respectively. For both types of energy sources, the proposed methods outperform the baseline.  Furthermore, the use of the optimal rotation for isotropic sources also offers an improvement over no rotation.
The $F_1$ score of it is shown in parenthesis, and the detailed results are shown in table \ref{table_noRotation} in the appendix. 

\begin{table}
	\centering
	\caption{Confusion matrices of detection statistics.}
	\label{table_ac1}\resizebox{0.8\columnwidth}{!}{%
	\begin{tabular}{|c|c|c|cccc|c|c|cccc|c|}
	\hline
    \multicolumn{2}{|c|}{}&\multicolumn{6}{c|}{isotropic sources}&\multicolumn{6}{c|}{skew sources}\\ \cline{3-14}
	\multicolumn{2}{|c|}{}& $b$&1 & 2 & 3 & 4 &$F_1$& $b$&1 & 2 & 3 & 4 &$F_1$\\ \hline
	\multirow{2}{*}{I}&2 &\multirow{2}{*}{2.25}&0 & {\bf 192}  & 0  &  0 & 0.98 &\multirow{2}{*}{1.57}&0& {\bf 192}  & 6  &  1  & \multirow{2}{*}{0.97}\\ \cline{4-7}\cline{10-13}
	&3 &&1& 0  & {\bf 193}  & 0  & (0.87)&&3& 0  & {\bf 196}  & 1  &\\ \hline\hline
\multirow{2}{*}{II}	&2&\multirow{2}{*}{0.8}&0& {\bf 200}  & 0  &  0&  \multirow{2}{*}{1.00} &\multirow{2}{*}{0.82}&0& {\bf 200}  & 0  &  0 & \multirow{2}{*}{1.00}\\ \cline{4-7}\cline{10-13}
	&3&&0& 0 & {\bf 200}  & 0 &   &&0& 0 & {\bf 200}  & 0   &\\ \hline\hline
	
\multirow{2}{*}{BL}&2&\multirow{2}{*}{0.42}&0& {\bf 106}  & 94  &  0& \multirow{2}{*}{0.68}&\multirow{2}{*}{0.575}&0& {\bf 121}  & 79  &  0 & \multirow{2}{*}{0.62}\\ \cline{4-7}\cline{10-13}
	 &3&&0& 30& {\bf 170}  & 0 &   &&0& 71 & {\bf 129}  & 0   &\\ \hline
    \end{tabular}}
    \vspace{-.1in}
\end{table}

\vspace{-0.1in}
\section{Conclusions}

This paper has presented several statistical detectors to determine the number of sources under the assumption of unimodal matrix models for each source. We have computed the asymptotic distributions of a new variance ratio statistic and assessed its performance gain relative to a baseline scheme that does not exploit unimodality. Our methods have the $F_1$ score around 1, while the baseline method has the $F_1$ score around 0.65. We also exploit a  rotation method to improve performance.  Optimal rotation has further significantly improved the performance (the $F_1$ score from 0.87 to 0.98).

\section{Acknowledgement}

The work of Rui Zhang and Yao Xie are supported by NSF CAREER Award CCF-1650913, and NSF DMS-1938106, and DMS-1830210. Junting Chen was supported by the Shenzhen Institute of Artificial Intelligence and Robotics for Society (AIRS) under Grant No. AC01202005001.




\clearpage
\bibliographystyle{IEEEbib}
\bibliography{refs}
\clearpage
\section{Appendix}
\subsection{Proof of Theorem \ref{thm-statsAsym}}

\begin{proof}
	Let $V = \hat\sigma_1^2(\boldsymbol{Z})/\sigma^2$ and $W = \hat\sigma^4_2(\boldsymbol{Z})/\sigma^4$. 
	By central limit theorem and  Slutsky's theorem \cite{casella2002statistical}, we have
	\begin{align*}
		\sqrt{L}\left(\left(\begin{matrix}
			V\\
			W
		\end{matrix}\right) - \left(\begin{matrix} 1\\ 1\end{matrix}\right)\right)
			\stackrel{d}{\rightarrow} \mathcal N(0, \Sigma),
	\end{align*}
	where 
\[
		\Sigma = \begin{bmatrix}
			\frac{2}{c} & \frac{4}{c}\\
			\\
			\frac{4}{c} & 2+\frac{12}{c}
		\end{bmatrix}.
\]
	Define $f(V,W) = \sqrt{W}/V$, and let $\nabla f$ denotes the gradient of $f$,
\[
		\nabla f(V,W) = (-\sqrt{W}/V^2 ,W^{-\frac{1}{2}}/(2V) )^\top
		.
\]
	Therefore, by delta method \cite{casella2002statistical}, we have
	\begin{align*}
		\sqrt{L}(f(V,W) - 1)\stackrel{d}{\rightarrow} \mathcal N(0, \Sigma_f),
	\end{align*}
	where
	\begin{align*}
		\Sigma_f = \nabla f(1,1)^\top\Sigma\nabla f(1,1) = \frac{c+2}{2c}.
	\end{align*}
\end{proof}

\subsection{Proof of Theorem \ref{thm-varRed}}

\begin{proof}
	Let $V = \hat\sigma_1^2(\boldsymbol{Z})/\sigma^2$ and $\tilde W = \hat\sigma^4_2(\boldsymbol{X})/\sigma^4$.
	With similar approach in the proof of theorem \ref{thm-statsAsym}, we have
	\begin{align*}
		\sqrt{L}\left(\left(\begin{matrix}
			V\\
			\tilde W
		\end{matrix}\right) - \left(\begin{matrix} 1\\ 1\end{matrix}\right)\right)
			\stackrel{d}{\rightarrow} \mathcal N(0, \tilde\Sigma),
	\end{align*}
	where 
\[\tilde\Sigma = \left(\begin{matrix}
			\frac{2}{c} & 0\\
			\\
			0 			& 2+\frac{12}{c}
		\end{matrix}\right).
\]
Define $f(V,\tilde W) = \frac{\sqrt{\tilde W}}{V}$, then we have
	\begin{align*}
		\sqrt{L}(f(V,\tilde W) - 1)\stackrel{d}{\rightarrow} \mathcal N(0, \tilde\Sigma_f),
	\end{align*}
	where
	\[	\tilde\Sigma_f = \nabla f(1,1)^\top\tilde\Sigma\nabla f(1,1) = \frac{c+10}{2c}.
\]
\end{proof}
\begin{table}[H]
	\centering
	\caption{Confusion matrix of variance ratio statistics without rotation.}
	\label{table_noRotation}
	\begin{tabular}{|c|cccc|c|}
	\hline
	$b=2.25$ \text{(no rotation)}	& 1 & 2 & 3 & 4& $F_1$ \\ \cline{1-6}
	$K^*=2$ &5& {\bf 156}  & 0  &  0 & \multirow{2}{*}{0.87}\\ \cline{1-5}
	$K^*=3$ &0& 18  & {\bf 171}  & 0 & \\ \hline


	\end{tabular}
\end{table}

\end{document}